\pdfoutput=1 
\newcommand*{\ATLASLATEXPATH}{}
\documentclass[cernpreprint,texlive=2016,txfonts,USenglish,texmf]{\ATLASLATEXPATH atlasdoc}
\usepackage{\ATLASLATEXPATH atlaspackage}
\usepackage{\ATLASLATEXPATH atlasbiblatex}

\usepackage{\ATLASLATEXPATH atlascontribute}

\usepackage{\ATLASLATEXPATH atlasphysics}
\graphicspath{{}{}}

\usepackage{MonoH_Paper_2016-defs}

\usepackage{multirow}
\usepackage{verbatim}
\usepackage{siunitx}
\AtlasTitle{Search for Dark Matter Produced in Association with a
            Higgs Boson Decaying to $b\bar b$ \\
            using 36 fb$^{-1}$ of $pp$ collisions at $\sqrt s=13$ TeV\\
            with the ATLAS Detector}

\author{The ATLAS Collaboration}

\AtlasRefCode{EXOT-2016-25}

\PreprintIdNumber{CERN-EP-2017-117}

\AtlasJournalRef{Phys. Rev. Lett. 119 (2017) 181804}
\AtlasDOI{10.1103/PhysRevLett.119.181804}

\AtlasAbstract{%
Several extensions of the Standard Model
predict associated production of dark-matter particles
with a Higgs boson. Such processes are searched for in final states
with missing transverse momentum and a Higgs boson decaying to a
$b\bar b$ pair with the ATLAS detector using 36.1 fb$^{-1}$ of $pp$
collisions at a center-of-mass energy of 13 TeV at the LHC.  The
observed data are in agreement with the Standard Model predictions and limits are
placed on the associated production of dark-matter particles and a
Higgs boson.  }

\begin{document}

\maketitle

%\tableofcontents

% List of contributors - print here or after the Bibliography.
%\PrintAtlasContribute{0.30}
%\clearpage

%%-------------------------------------------------------------------------------
%\section{Introduction \textcolor{blue}{John}}
%\label{sec:intro}
%%-------------------------------------------------------------------------------

One of the central open questions in physics today is the 
nature of dark matter (DM) that comprises most of the matter in
the universe~\cite{Bertone:2004pz}. A compelling candidate for DM is a
stable electrically neutral particle $\chi$ whose non-gravitational
interactions with Standard Model (SM) particles are weak. This
extension of the SM could be detectable at the scale of electroweak symmetry
breaking~\cite{Jungman:1995df} and accommodate the observed DM relic
density~\cite{Steigman:1984ac,Scherrer:1985zt}. Many models predict
detectable production rates of such DM particles at the Large Hadron
Collider (LHC)~\cite{Abercrombie:2015wmb}.

Most collider-based searches for DM rely on the signature of missing
transverse momentum\footnote{ATLAS uses a right-handed coordinate
  system with its origin at the nominal interaction point (IP) in the
  center of the detector and the $z$-axis along the beam pipe. The
  $x$-axis points to the center of the LHC ring, and the
  $y$-axis points upward. Cylindrical coordinates $(r,\phi)$ are used
  in the transverse plane, $\phi$ being the azimuthal angle around the
  $z$-axis. The pseudorapidity is defined in terms of the polar angle
  $\theta$ as $\eta=-\ln\tan(\theta/2)$. The distance between two
  objects in $\eta$--$\phi$ space is $\Delta R = \sqrt{(\Delta \eta)^2
    + (\Delta \phi)^2}$. Transverse momentum is defined by $\pt =
  p\sin{\theta}$.} \met from DM particles recoiling against one SM
particle $X$ radiated off the initial state, denoted by the
``$X+\met$'' signature.  LHC experiments have searched for this
$X+\met$ signature, where $X$ is a light quark or
gluon~\cite{Aad:2015zva, Khachatryan:2014rra,Aaboud:2016tnv}, a $b$-
or $t$-quark~\cite{Aad:2014vea, Khachatryan:2014uma,
  Khachatryan:2015nua}, a
photon~\cite{Aad:2014tda,Chatrchyan:2012tea,Khachatryan:2014rwa,Khachatryan:2014tva,Aaboud:2016uro},
or a $W$ or $Z$
boson~\cite{ATLAS:2014wra,Aad:2014vka,Aad:2013oja,Khachatryan:2015bbl}. The
discovery of the Higgs boson $h$~\cite{Aad:2012tfa,Chatrchyan:2012ufa}
opens a new opportunity 
through the $h + \met$
signature~\cite{Petrov:2013nia,Carpenter:2013xra,Berlin:2014cfa}.
Because $h$ radiation off the initial state is
Yukawa-suppressed, the $h + \met$ process represents a direct probe
of the hard interaction involving DM particles.

This Letter presents a search for DM in association with a Higgs boson 
decaying to a pair of $b$-quarks, $h\to b\bar b$, with a branching ratio $\br=57\%$~\cite{deFlorian:2016spz}, using 36.1~\ifb\ of
$pp$ collisions at $\rts = 13~\TeV$ recorded with the ATLAS detector~\cite{Aad:2008zzm,Capeans:1291633} in Run~2 of the LHC in
2015 and 2016. This search substantially extends the sensitivity
relative to previous results at 8~\TeV~\cite{Aad:2015dva,Aad:2015yga} and
13~\TeV~\cite{bib:monoh2015,Aaboud:2017uak,Sirunyan:2017hnk} in the $h \to b\bar b$ and $h \to
\gamma\gamma$ channels. 

A Type-II two-Higgs-doublet model (2HDM) with an
additional U(1)$_{\zprime}$ gauge symmetry yielding an additional massive
\zprime boson provides an $h + \met$ signature~\cite{Berlin:2014cfa} used for the optimization
of the search and its interpretation. This model results
in five physical Higgs bosons: a light scalar $h$ identified with the SM
Higgs boson in the alignment limit~\cite{Branco:2011iw}, a heavy
scalar $H$, a pseudoscalar $A$, and two charged scalars $H^\pm$.  The
$h+\textrm{DM}$ signal in this \zhdm model is produced through $pp\to Z'\to Ah$, 
where $A$ decays to $\chi\bar{\chi}$ with a large 
$\br$.  Relevant model parameters are: the ratio of the
vacuum expectation values of the two Higgs fields coupling to the
up-type and down-type quarks $\tan\beta$, the \zprime gauge coupling
$g_{\zprime}$, and the masses $m_{\zprime}$, $m_A$, and $m_\chi$.
The results are also generically interpreted in terms of
the production cross-section of non-SM events with large \met and a Higgs boson 
without extra model assumptions.

%%-------------------------------------------------------------------------------
%\section{Data and simulation samples \textcolor{blue}{John}}
%\label{sec:dataMC}
%%-------------------------------------------------------------------------------

Monte Carlo (MC) event generators were used to simulate the $h+\textrm{DM}$ signal
and all SM background processes, except the multijet
background, which was evaluated using data. All MC event samples were processed through a detailed simulation of
the ATLAS detector~\cite{Aad:2010ah} based on
\textsc{Geant4}~\cite{Geant4}, and contributions from
additional $pp$ interactions~(pileup) were simulated using \textsc{Pythia} 8.186~\cite{pythia8} and the MSTW2008LO  parton
distribution function (PDF) set~\cite{PDF-MSTW2008}.

Signal samples for the $pp\to Z'\to Ah \to \chi\bar{\chi}b\bar b$
process were generated at leading order using
\textsc{MadGraph}\_aMC@NLO
2.2.3~\cite{Abercrombie:2015wmb,Alwall:2014hca} interfaced to
\textsc{Pythia} 8.186, using the NNPDF3.0 PDF set~\cite{Ball:2014uwa}.  Samples were
generated in the $(m_{Z'},m_A)$ plane for $0.2~\TeV<m_{Z'}<3~\TeV$ and
$0.2~\TeV<m_A<0.8~\TeV$ with $m_{\chi}=100~\GeV$, $\tan\beta = 1$,
$g_{\zprime} = 0.8$, $m_H=m_{H^\pm}=300~\GeV$~\cite{Abercrombie:2015wmb}.
%, and assuming $\br(A\to \chi\bar{\chi})=100\%$~\cite{Abercrombie:2015wmb}.

Backgrounds from top quark pair production and single top quark
production were generated at next-to-leading order (NLO) in quantum chromodynamics (QCD) with
\textsc{Powheg-box}~\cite{Nason:2004rx,Frixione:2007vw,Powheg,Alioli:2009je,Re:2010bp} using CT10 PDFs~\cite{Gao:2013xoa}, where the parton
shower was simulated with \textsc{Pythia}
6.428~\cite{Sjostrand:2006za}.  The $t\bar{t}$ samples are normalized
using calculations at next-to-next-to-leading order (NNLO) in QCD including
next-to-next-to-leading logarithmic corrections for soft-gluon
radiation~\cite{Czakon:2013goa}.  The single-top-quark processes are
normalized with cross-sections at NLO in QCD~\cite{Bordes:1994ki, PhysRevD.56.5919, Stelzer:1998ni, Smith:1996ij,Kidonakis:2013zqaHacked}.
Background processes involving a vector boson $V=W,Z$ decaying leptonically in association with jets, \vjets, were simulated with \textsc{Sherpa} 2.2.1~\cite{Gleisberg:2008ta} including mass effects for $b$- and $c$-quarks and using NNPDF3.0 PDFs. The perturbative calculations for \vjets were performed at NLO for up to two partons and at leading order for up to four partons~\cite{Gleisberg:2008fv,Cascioli:2011va}, and matched to the parton shower~\cite{Schumann:2007mg} using the ME+PS@NLO prescription from Ref.~\cite{Hoeche:2012yf}. The normalizations are determined at NNLO in QCD~\cite{Melnikov:2006kv}.
Diboson processes ($VV$) were simulated at NLO in QCD with \textsc{Sherpa} 2.1.1 and CT10 PDFs.
Backgrounds from associated $Vh$ production were generated with
\textsc{Pythia} 8.186 using NNPDF3.0 PDFs for $qq\to Vh$, and
\textsc{Powheg} interfaced to \textsc{Pythia} 8.186 using CT10 PDFs
for $gg\to Vh$.

%%-------------------------------------------------------------------------------
%\section{Object reconstruction \textcolor{blue}{John}}
%\label{sec:objects}
%%-------------------------------------------------------------------------------

Events are selected by an \met trigger based on calorimeter
information~\cite{Aaboud:2016leb}.  
Its threshold was $110~\GeV$ for most of the data taking period, and lower in the first third.  
Events are required to have at least one $pp$ collision
vertex reconstructed from at least two inner detector (ID) tracks
with $\pt^{\mathrm{track}} > 0.4~\GeV$.  The primary vertex~(PV) for
each event is the vertex with the highest
$\sum(\pt^{\mathrm{track}})^2$.

Reconstruction of muons ($\mu$) incorporates tracks or track segments
found in the muon spectrometer and matched ID tracks.  Identified
muons must satisfy the ``loose'' quality
criteria~\cite{Aad:2016jkr} and have $|\eta|<2.7$.  Electrons~($e$) are
reconstructed by matching an ID track to a cluster of energy in the
calorimeter.  Electron candidates are identified through a likelihood-based
method~\cite{Aaboud:2016vfy} and must satisfy the
``loose'' operating point and be within $|\eta|<2.47$.  Muon and
electron candidates must have $\pt> 7~\GeV$ and are required to be
isolated by limiting the sum of \pt for tracks within a cone in
$\Delta R$ around the lepton direction, as in
Ref.~\cite{bib:monoh2015}.

Jets reconstructed from three-dimensional clusters of calorimeter cells~\cite{Aad:2016upy} with the \antikt algorithm~\cite{Cacciari:2008gp} are used to identify the $h\to b\bar{b}$ decay. For small to moderate $h$ momenta, the decay products can be resolved using jets with a radius parameter $R=0.4$ (small-$R$ jets or $j$).  The decay products of high-momenta $h$ become collimated and are reconstructed using a single jet with $R=1.0$ (large-$R$ jet or $J$).
Small-$R$ jets with $|\eta|<2.5$ must
satisfy $\pt > 20~\GeV$ and are called ``central'', while those with
$2.5<|\eta|<4.5$ must have $\pt > 30~\GeV$ and are called
``forward''. Small-$R$ jets are corrected for pileup~\cite{Cacciari:2008gn}, and central small-$R$ jets with
$20~\GeV <\pt<60~\GeV$ and $|\eta|<2.4$ are additionally required to be identified as originating from the PV using associated tracks~\cite{ATLAS:2014cva}.  Small-$R$
jets closer than $\Delta R=0.2$ to an electron candidate
are rejected.  
Large-$R$ jets are reconstructed independently of small-$R$ jets and trimmed~\cite{Krohn2010,ATLAS-CONF-2016-039} to reduce the effects of
pileup and the underlying event. 
Furthermore, large-$R$ jets must fulfill
$\pt>200~\GeV$ and $|\eta|<2.0$.  To improve the resolution and
minimize uncertainties, the mass of large-$R$ jets is determined by
the resolution-weighted mean of the mass measured using 
calorimeter information alone and the track-assisted
jet mass~\cite{ATLAS-CONF-2016-035}. 
The latter is obtained by scaling
the mass determined using ID tracks alone by the ratio of jet \pt measured in the
calorimeter and in the ID.

Multivariate algorithms are used to identify jets containing
$b$-hadrons ($b$-tagging), which are expected in $h \to b\bar b$
decays~\cite{ATLAS-CONF-2016-039,ATL-PHYS-PUB-2016-012}. These
algorithms are applied directly to small-$R$ jets, while for large-$R$
jets they are applied to track-jets matched to large-$R$
jets. Track-jets are reconstructed from ID tracks matched to the PV
using the \antikt algorithm with $R=0.2$, and must fulfill $\pt>
10~\GeV$ and $|\eta|<2.5$.

The \vecmet observable is calculated as the negative of the vector sum
of the transverse momenta of $e$, $\mu$, and jet candidates in the event. The transverse momenta
not associated with any $e$, $\mu$, or jet candidates are accounted for using ID
tracks~\cite{ATL-PHYS-PUB-2015-023,Aad:2012re}. Similarly, 
$\vecmpt$ is defined as the negative of the
vector sum of the transverse momenta of tracks with $\pt > 0.5~\GeV$ associated with the PV
and within $|\eta|<2.5$.

%%-------------------------------------------------------------------------------
%\section{Event selection \textcolor{green}{Oleg}}
%\label{sec:selection}
%%-------------------------------------------------------------------------------

The signal is characterized by high \met, no isolated leptons, and an
invariant mass of the $h$ candidate \mh compatible with the observed
Higgs boson mass of 125~\GeV~\cite{Aad:2015zhl}. In the signal
region (SR) described below, the dominant backgrounds from
$Z(\nu\nu)+\mathrm{jets}$, $W+\mathrm{jets}$, and \ttbar production contribute respectively
 30--60\%, 10--25\%, and 15--50\% of the total
background, depending on \met and the $b$-tag multiplicity. The
models for \vjets and \ttbar are constrained using two control
regions (CR): the single-muon control region (\onemucr) is designed to
constrain the \ttbar and \wjets backgrounds, while the two-lepton
control region (\twolepcr) constrains the \zjets background
contribution.

The SR requires $\met>150\,\gev$, and no isolated $e$ or $\mu$. 
The multijet background contributes due to mismeasured jet momenta. To suppress it, additional selections are required: $\min\left[\Delta\phi\left(\vecmet,\vec p_\mathrm{T}^j\right)\right]>\pi/9$ for the three highest-\pt~(leading) \smallR jets,  $\Delta\phi\left(\vecmet,\vecmpt\right)<\pi/2$, and $\mpt>30\,\gev$ for events with fewer than two central $b$-tagged \smallR jets. The requirements using \mpt also reduce non-collision backgrounds.

In the ``resolved'' regime, defined by $\met<500~\GeV$, the $h$ candidate is reconstructed from two leading $b$-tagged central \smallR jets, or, if only one $b$-tag is present in the event, from the $b$-tagged central \smallR jet and the leading non-$b$-tagged central \smallR jet. At least one of the jets comprising the $h$ candidate must satisfy $\pt>45$~GeV. 
A separation in $\Delta\phi$ between the $h$ candidate and \vecmet of more than $2\pi/3$ is required following the back-to-back configuration of the Higgs boson recoiling against DM. 
To improve the trigger efficiency modeling, events are retained only if the
scalar sum $H_\mathrm{T}$ of the \pt~of the two (three) leading jets fulfills
$H_{\mathrm{T},2j}>120~\gev$ ($H_{\mathrm{T},3j}>150~\gev$)
if two (more than two) central jets are present. Further optimization
of the event selection described below provides an additional
background reduction of up to 60\% relative to
Ref.~\cite{bib:monoh2015}, for a small signal loss.
Events with a hadronic $\tau$-lepton candidate, identified either by an
algorithm based on a boosted decision tree~\cite{Aad:2015unr} or as
\smallR jets containing one to four tracks within the jet core and
$\Delta\phi\left(\vecmet,\vec p_\mathrm{T}^j\right)<\pi/8$, are
rejected to reduce the \ttbar background, which can enter the SR if at
least one top quark decays as $t \to Wb\to \tau\nu b$.  
This background is further reduced by removing events with more than two
$b$-tagged central jets, which typically happens for \ttbar events with $t \to
Wb\to c s b$ decays. 
%This background is further reduced by removing events with more than two
%$b$-tagged central jets, which essentially consist of \ttbar events with 
%one $t \to Wb\to c s b$ decay. 
Since most of the hadronic activity in a
signal event is expected from the $h\to b\bar b$ decay, the scalar sum
of the \pt of the two jets forming the $h$ candidate and, if present, the
highest-\pt additional jet must be larger than $0.63\times
H_{\mathrm{T,all\,jets}}$. Finally, $\Delta R\left(\vec
p^{\,j_1}_h,\vec p^{\,j_2}_h\right)<1.8$ is required for the two jets
forming the $h$ candidate.

In the ``merged'' regime, defined by $\met>500~\GeV$, the leading \largeR
jet represents the $h$ candidate. Further 
selection optimization reduces backgrounds, primarily \ttbar production, by up to 30\% relative to
Ref.~\cite{bib:monoh2015}, for a small signal loss: events containing
$\tau$-lepton candidates with $\Delta R\left(\vec
p^{\,\tau},\vec
p^{\,J}\right)>1.0$ are vetoed; no $b$-tagged
central \smallR jets with $\Delta R\left(\vec
p^{\,j,b\textrm{-tag}},\vec p^{\,J}\right)>1.0$ are allowed in
the event; and the scalar sum of \pt of the \smallR jets with
$\Delta R\left(\vec p^{\,j},\vec
p^{\,J}\right)>1.0$ is required to be smaller than $0.57$ times that sum added to $\pt^J$. 

The resolution in \mh is improved using muons associated with small-$R$ jets in the resolved regime or with track-jets matched to large-$R$ jets in the merged regime~\cite{Aad:2014xzb,ATLAS-CONF-2016-039}.

%%-------------------------------------------------------------------------------
%\section{Background estimation \textcolor{green}{CRs: Oleg}
%\textcolor{blue}{Multijet background: John}
%}
%\label{sec:bgr}
%%-------------------------------------------------------------------------------

The event selection in the \onemucr is identical to the SR, except that exactly one isolated $\mu$ candidate with $\pt^\mu>27~\GeV$ is required, and that $\vec{p}_{\mathrm T}^{\,\mu}$ is added to \vecmet to mimic the behavior of events contaminating the SR when the charged lepton is not detected. 

Events in the \twolepcr are collected using a single-$e$ or
single-$\mu$ trigger, and selected by requiring one pair of isolated
$e$ or $\mu$, one of which must have $\pt^\ell>27~\GeV$. Events with a
$Z$ boson candidate are retained, identified as having $83~\GeV <
m_{ee} < 99~\gev$ or $71~\gev < m_{\mu\mu} < 106~\gev$ with an
opposite-charge requirement in the $\mu\mu$ case. In addition, a
measure of the \met significance given by the ratio of the \met to the
square root of the scalar sum of \pt of all leptons and small-$R$ jets in the
event must be less than $3.5\,\gev^{1/2}$. This requirement
separates $Z(\ell\ell)$+jets processes from \ttbar production, as \met
originates from finite detector resolution for the former and mainly from
neutrinos for the latter. To mimic $Z\to\nu\nu$
decays in the SR, the \vecmet is set to the
$\vec{p}_\mathrm{T}$ of the dilepton system, which is then ignored in the
subsequent analysis. All other selection requirements are identical between
the \twolepcr and the SR.

Subdominant backgrounds, including diboson, $Vh$, single top quark, and multijet
production, contribute less than 10\% of the total background in the SR. Multijet production is negligible for $\met>350~\GeV$. Its \mh distribution is determined from data in a
dedicated  multijet-enriched sideband, defined by inverting the
$\min\left[\Delta\phi\left(\vecmet,\vec
  p_\mathrm{T}^j\right)\right]$ requirement.

%%-------------------------------------------------------------------------------
%\section{Systematic uncertainties \textcolor{green}{Oleg}}
%\label{sec:syst}
%%-------------------------------------------------------------------------------
Dominant sources of experimental systematic uncertainty arise from the number of background MC events, the calibration of the $b$-tagging efficiency and integrated luminosity, as well as the scale and resolution of the energy and the mass of jets. Uncertainties associated with the $\tau$ vetoes are found to be negligible. Dominant sources of theoretical systematic uncertainty originate from the modeling of the signal and background processes such as \ttbar, \vjets, $Vh$, diboson, and multijet production. The few relevant changes in the estimation of systematic uncertainties relative to Ref.~\cite{bib:monoh2015} encompass: the improved calibrations of the $b$-tagging efficiency using \ttbar events~\cite{ATLAS-CONF-2016-039,ATL-PHYS-PUB-2016-012} as well as of the jet energy and mass scales using various {\em in situ} methods~\cite{PERF-2016-04,ATLAS-CONF-2016-035}; the reduced uncertainty from the new jet-mass observable~\cite{ATLAS-CONF-2016-039,ATLAS-CONF-2016-035}; and the uncertainty of 3.4\% on the integrated luminosity of data collected in 2016. 
Table~\ref{tab:systs} quantifies dominant sources of uncertainty after the fit to data assuming three representative \zhdm-scenarios. This search is statistically limited for $\met\gtrsim300~\GeV$.

\begin{table}
\caption{
\label{tab:systs}
Dominant sources of uncertainty for three representative
\zhdm scenarios after the fit to data: (a)~with $(m_{Z'},m_A)=(0.6~\TeV,0.3~\TeV)$, (b)~with $(m_{Z'},m_A)=(1.4~\TeV,0.6~\TeV)$, and (c)~with $(m_{Z'},m_A)=(2.6~\TeV,0.3~\TeV)$. The effect is expressed as the fractional uncertainty
on the signal yield. %, assuming total cross-sections of (a)~452~fb, (b)~\sxsec~fb, and (c)~2.03~fb. 
Total is the quadrature sum of statistical and total systematic uncertainties.
The impact of the luminosity uncertainty, which does not affect backgrounds with free normalizations, varies due to the changing background composition with increasing \met.
} \centering
%\vspace{-3mm}
\begin{tabular}{lrrr}
\hline
\hline
\multirow{2}{*}{Source of uncert.} & \multicolumn{3}{c}{Impact [\%]} \\
                                   & (a) & (b) & (c) \\
\hline
\vjets modeling                       & 5.0  & 5.7   & 8.2 \\
$t\bar{t}$, single-$t$ modeling       & 3.2  & 3.0   & 3.9 \\
SM $Vh(b\bar b)$ norm.                & 2.2  & 6.9   & 6.9 \\
Signal modeling                       & 3.9  & 2.9   & 2.1 \\
MC statistics                         & 4.9 & 11\phdo  & 22\phdo     \\
Luminosity                            & 3.2  & 4.5   & 5.4 \\
$b$-tagging, track-jets               & 1.4 & 11\phdo  & 17\phdo     \\
$b$-tagging, calo jets                & 5.0 & 3.4   & 4.7 \\
Jets with $R=0.4$                     & 1.7  & 3.8   & 2.1 \\
Jets with $R=1.0$                     & $<$0.1 & 1.2 & 4.7  \\
\hline
Total syst. uncert.                   & 10\phdo & 21\phdo  & 36\phdo    \\
Statistical uncert.                   & 6\phdo & 38\phdo  & 62\phdo     \\
Total uncert.                         & 12\phdo  & 43\phdo   & 71\phdo     \\
\hline
\hline
\end{tabular}
\vspace{-5mm}
\end{table}

A fit to the \mh observable based on a binned likelihood approach~\cite{Moneta:2010pm,Verkerke:2003ir_} is used to search for a signal.
Systematic uncertainties are included in the likelihood function as nuisance parameters with Gaussian or log-normal constraints and profiled~\cite{Aad:2014xzb}. 
To account for changes in the background composition and to benefit from a higher signal sensitivity with increasing \met and $b$-tag multiplicity, the data are split into categories that are fit simultaneously. 
Eight categories are defined for the SR and each of the two CRs: four ranges in $\met/\GeV$ as $[150,200)$, $[200,350)$, $[350,500)$, and $[500,\infty)$, which are each split into two subregions with one and two $b$-tags. 
In the \onemucr, the electric charge of the $\mu$ is used to separate \ttbar from \vjets since the former provides an equal number of $\mu^+$ and $\mu^-$,  while a prevalence of $\mu^+$ is expected from the latter process due to PDFs~\cite{Kom:2010mv}. Only the total event yield is considered in the \twolepcr due to limited data statistics. The normalizations of \ttbar, $W\!+\!\textrm{HF}$, and $Z\!+\!\textrm{HF}$ processes are free parameters in the fit, where HF represents jets containing $b$- or $c$-quarks. 
In the SR, the contribution from \zjets is increased by about 50\% by the fit relative to theory predictions, staying within uncertainties, while \ttbar is reduced by up to 30\% at high \met.
% In the SR, the normalization of \zjets is increased by about 50\% by the fit, while \ttbar is reduced by up to 30\% at high \met. 
The normalizations of other backgrounds modeled using MC simulations are constrained to theory predictions within uncertainties, as detailed in Ref.~\cite{bib:monoh2015}.

The distributions of \mh for SR events with two $b$-tags provide the highest signal sensitivity and are shown in the four \met regions in Figure~\ref{fig:mbb}. No significant deviation from SM predictions is observed.

\begin{figure*}
\vspace{-2mm}
\centering
\includegraphics[width=0.45\textwidth]{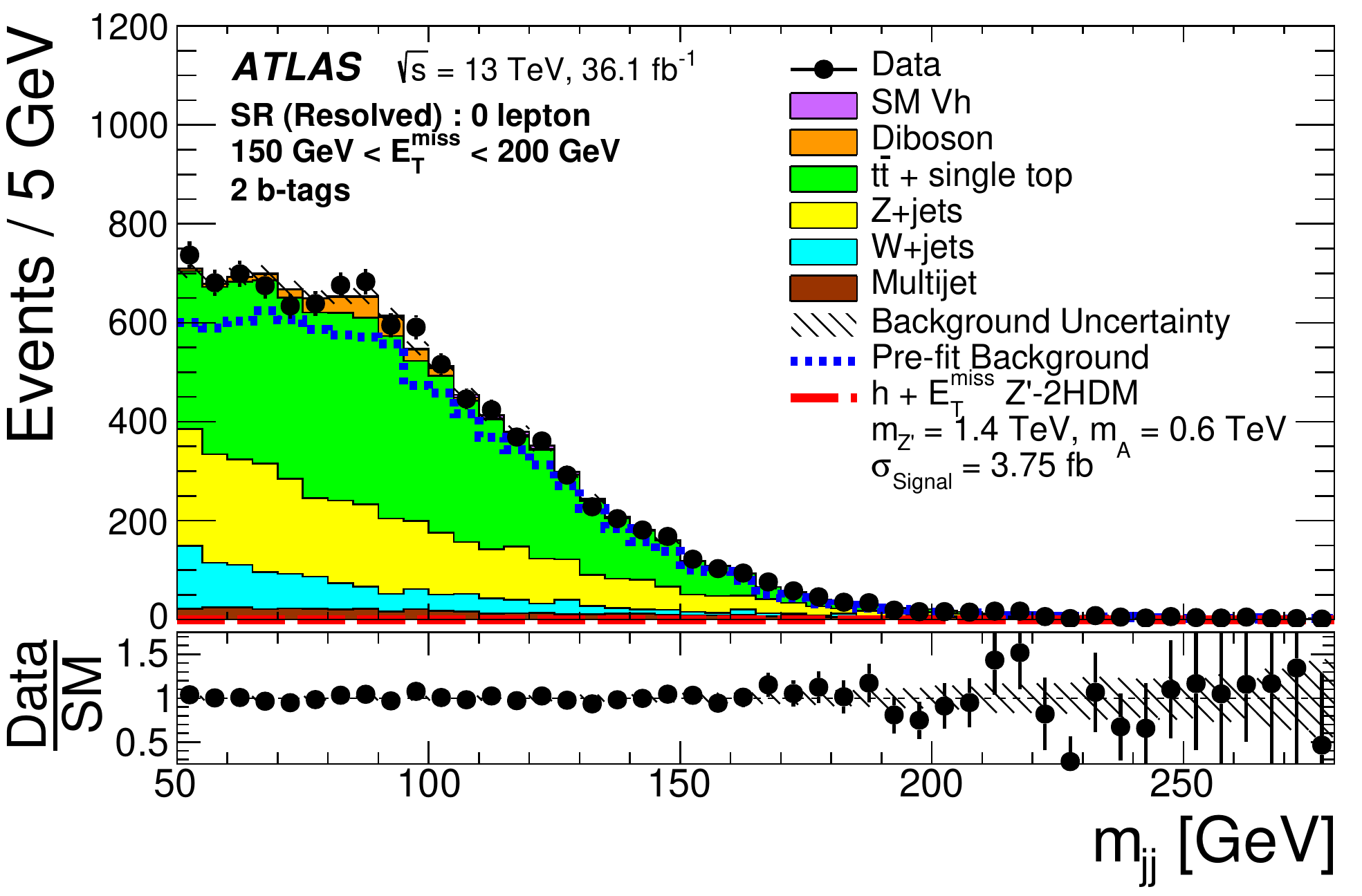}
\includegraphics[width=0.45\textwidth]{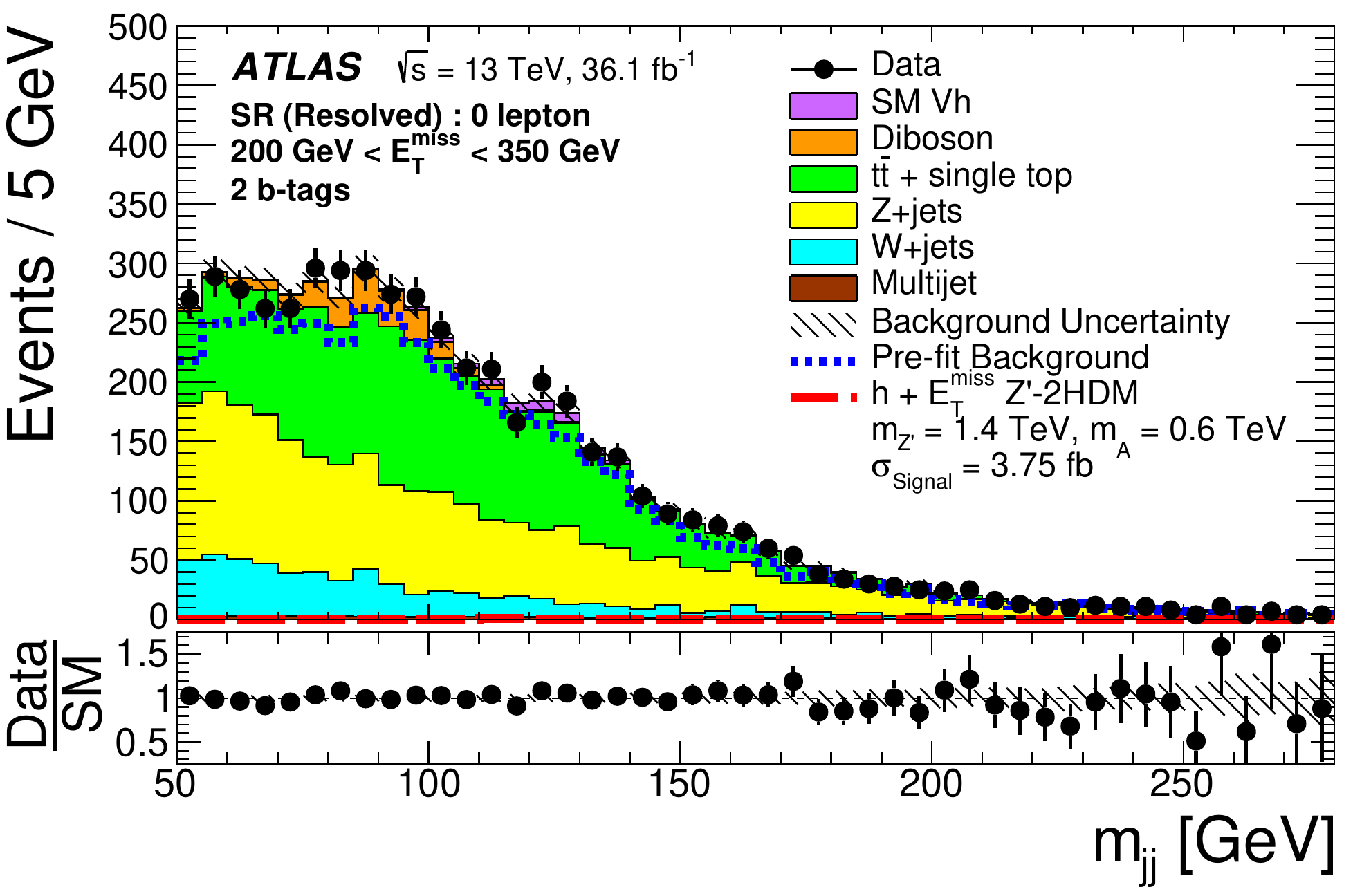}\quad
\includegraphics[width=0.45\textwidth]{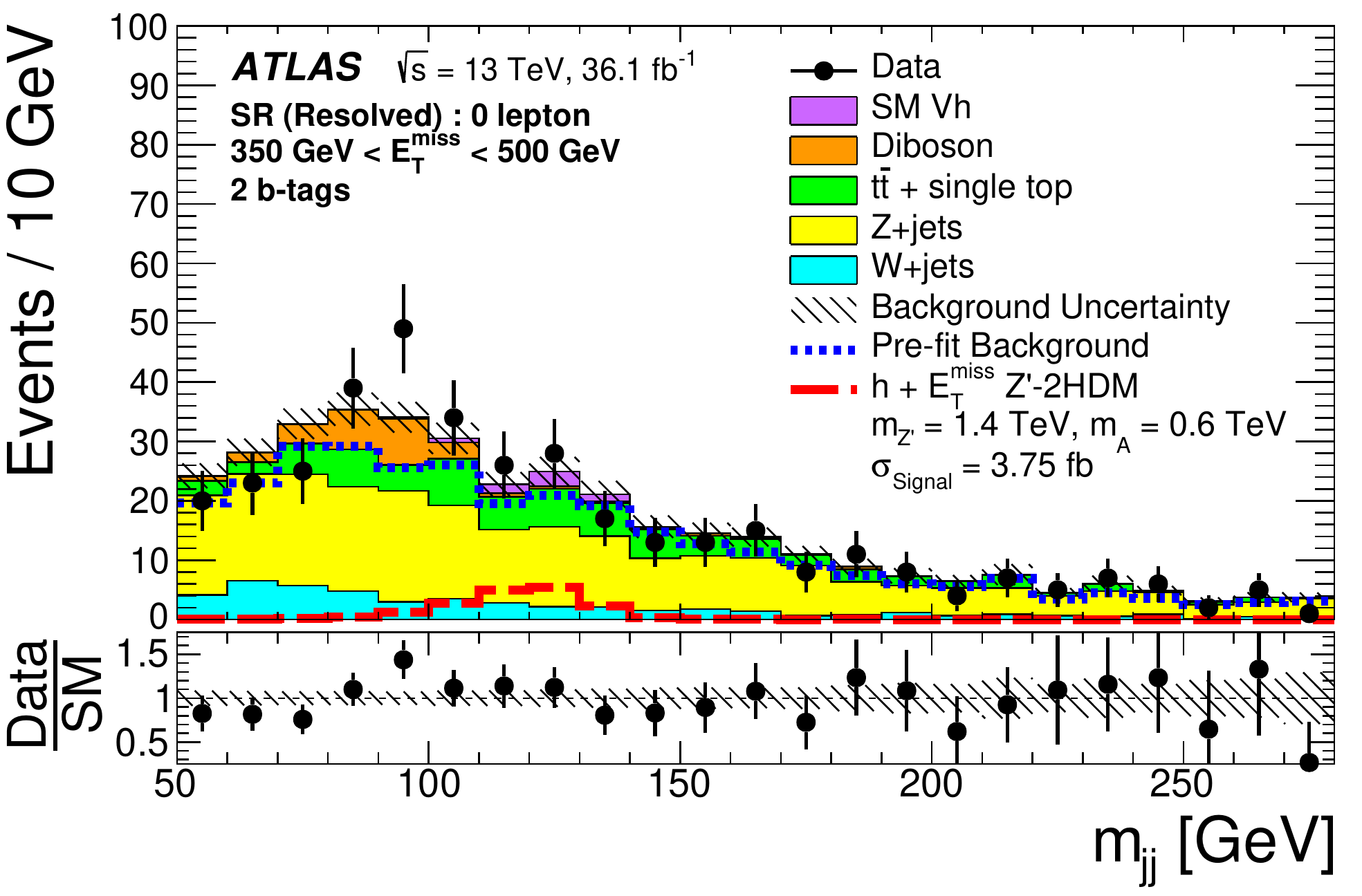}
\includegraphics[width=0.45\textwidth]{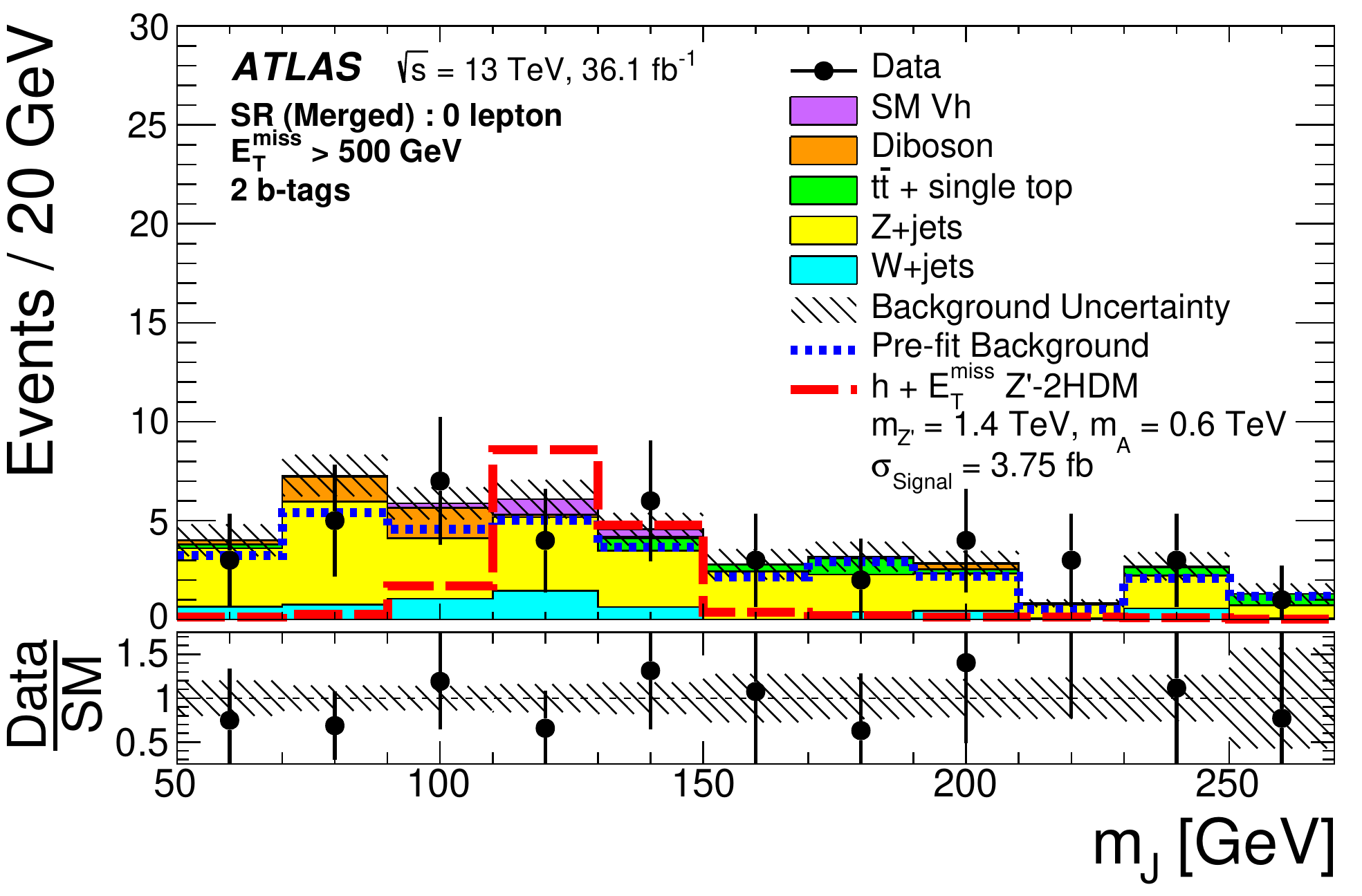}
\vspace{-3mm}
\caption{ Distributions of the invariant mass of the Higgs
  boson candidates $\mh=m_{jj},m_J$ with two $b$-tags in the SR for the four \met categories that are used as inputs to
  the fit. The upper panels show a comparison of data to the SM
  expectation before~(dashed lines) and after the fit~(solid histograms) with
  no signal included. The lower panels display the
  ratio of data to SM expectations after the fit, with its systematic
  uncertainty considering correlations between individual contributions indicated by the hatched band. The expected signal from a
  representative \zhdm model is also shown~(long-dashed line).
\vspace{-3mm}
\label{fig:mbb}
}
\end{figure*}

The results are interpreted as exclusion limits  at 95\% confidence level (CL) on the production cross-section of $h+\textrm{DM}$ events $\sighdm$ times $\br(h\to b\bar b)$ with the CL$_\mathrm{s}$ formalism~\cite{Read:2002hq} using a profile likelihood ratio~\cite{Cowan:2010js} as test statistic. Exclusion contours in the $(m_{\zprime},m_A)$ space in the \zhdm scenario are presented in Figure~\ref{fig:limits}, excluding $m_{\zprime}$ up to 2.6~\TeV\ and $m_A$ up to 0.6~\TeV, substantially extending previous limits~\cite{Aad:2015dva,Aad:2015yga,bib:monoh2015,Aaboud:2017uak,Sirunyan:2017hnk}.
Furthermore, upper limits on $\sighdm\times \br(h\to b\bar b)$ are provided under the minimal $h+\textrm{DM}$ model assumption that a Higgs boson is produced in a generic back-to-back configuration relative to \vecmet from DM particles. For this, limits are set on $\sigvishdm\equiv \sighdm\times \br(h\to b\bar b)\times \acc\times\varepsilon$ of $h(b\bar b)+\textrm{DM}$ events per \met bin at detector level, after all SR selections except the requirements on $b$-tag multiplicity and \mh range as used in the fit. The $\acc\times\varepsilon$ term quantifies the probability for an event to be reconstructed in the same \met bin as generated and to pass all $\sigvishdm$ selections, where \acc represents the kinematic acceptance and $\varepsilon$ accounts for the experimental efficiency.
The results are  shown in Table~\ref{tab:mil}. To minimize the dependence on the \met distribution of a potential $h+\textrm{DM}$ signal, the standard fit approach is modified to analyze one \met range at a time in the SR. 
The \zhdm model is used to evaluate the dependence of
the $\sigvishdm$ limits and of $\acc\times\varepsilon$
on the event kinematics within a given \met bin. 
%, assuming a generic back-to-back topology of the Higgs boson and \vecmet. 
A range of $(m_{Z'},m_A)$ parameters that yield a sizable contribution of $\gtrsim10\%\times\sighdm\times \br(h\to b\bar b)$ in a given \met bin is considered. Corresponding variations of 25\% (70\%) in the expected limits and of 50\% (25\%) in $\acc\times\varepsilon$ are found in the resolved (merged) regime. Table~\ref{tab:mil} quotes the least stringent limit and the lowest $\acc\times\varepsilon$ value in a given \met bin after rounding. The limits are valid for $p_{\mathrm{T},h}\lesssim1.5~\TeV$.

\begin{figure}
\vspace{-2mm}
\centering
\includegraphics[width=0.43\textwidth]{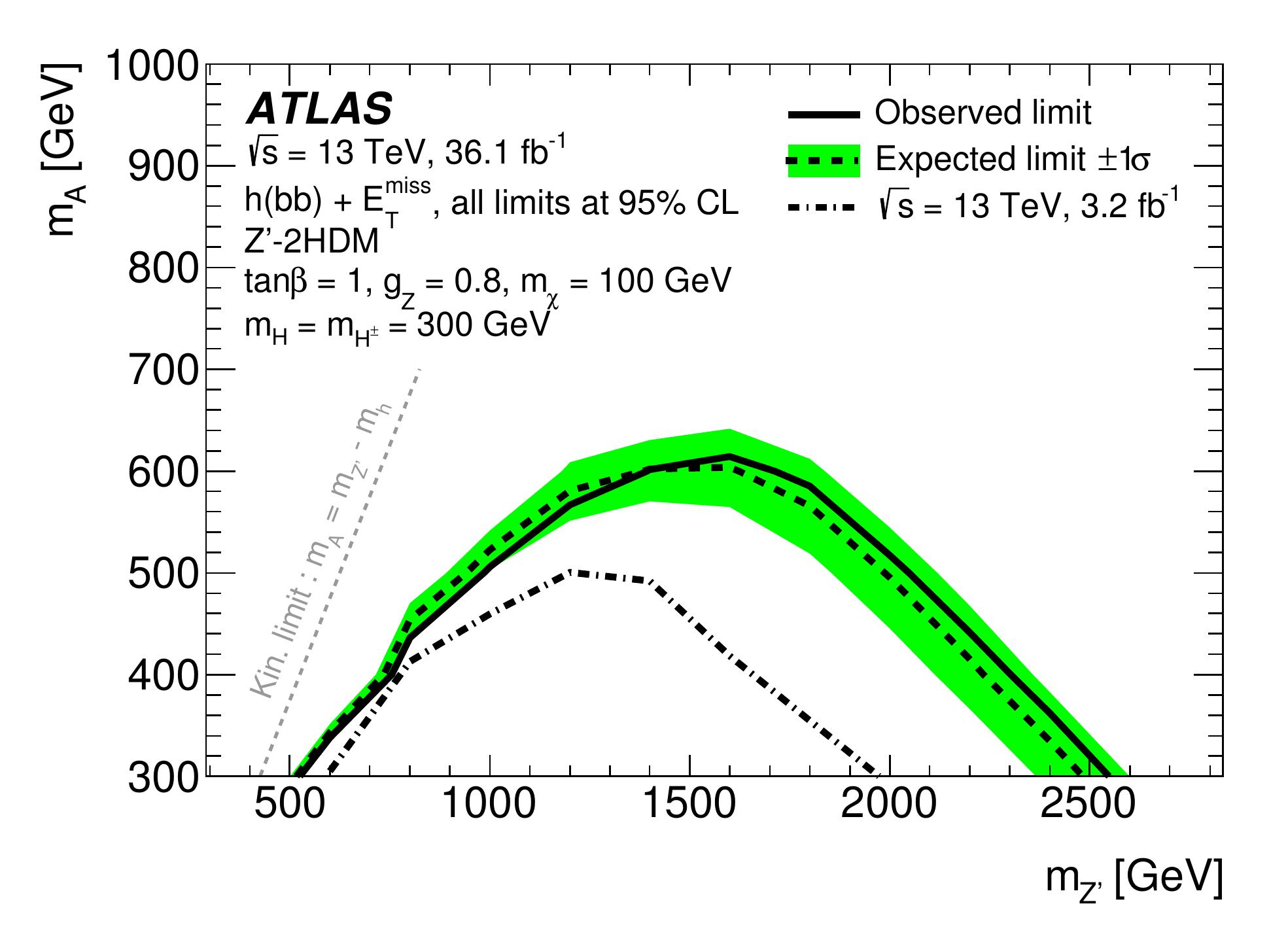}
\vspace{-3mm}
\caption{
Exclusion contours for the \zhdm scenario in the $(m_{\zprime},m_{A})$ plane for $\tan{\beta} = 1$, $g_{\zprime}=0.8$, and $m_{\chi}=100\,\gev$. The observed limits (solid line) are consistent with the expectation under the SM-only hypothesis (dashed line) within uncertainties (filled band). Observed limits from previous ATLAS results at $\sqrt s =13~\TeV$ (dash-dotted line)~\cite{bib:monoh2015} are also shown.
\label{fig:limits}
\vspace{-3mm}
}
\end{figure}

\renewcommand{\arraystretch}{1.15}
\begin{table}
\begin{center}
\caption{
Observed (obs) and expected (exp) upper limits at 95\% CL on $\sigvishdm\equiv \sighdm\times \br(h\to b\bar b)\times \acc\times\varepsilon$ of $h(b\bar b)+\textrm{DM}$ events. Also shown are the acceptance $\times$ efficiency ($\acc\times\varepsilon$) probabilities to reconstruct and select an event in the same \met bin as generated.
\label{tab:mil}
}
%\footnotesize
%\vspace{-3mm}
\begin{tabular}{lccc}
\hline \hline
Range in & $\sigvishdm^\mathrm{obs}$ &   $\sigvishdm^\mathrm{exp}$ & $\acc\times\varepsilon$ \\
$\met$ [\GeV] & [fb] & [fb] & [\%] \\
\hline
$[150,200)$    & 19.1 & $18.3^{+7.2}_{-5.1}$ & 15 \\
$[200,350)$    & 13.1 & $10.5^{+4.1}_{-2.9}$ & 35 \\
$[350,500)$    & \phantom{1}2.4 & $\phantom{1}1.7^{+0.7}_{-0.5}$ & 40  \\
$[500,\infty)$ & \phantom{1}1.7 & $\phantom{1}1.8^{+0.7}_{-0.5}$ & 55\vspace{0.7mm} \\
\hline \hline
\end{tabular}
\end{center}
\vspace{-5mm}
\end{table}
\renewcommand{\arraystretch}{1}

% All figures and tables should appear before the summary and conclusion.
% The package placeins provides the macro \FloatBarrier to achieve this.
\FloatBarrier

%%-------------------------------------------------------------------------------
%\section{Conclusion}
%\label{sec:conclusion}
%%-------------------------------------------------------------------------------
In summary, a search for DM produced in association with a Higgs boson
in final states with \met and a $b\bar b$ pair from the $h \to b\bar
b$ decay was conducted using $36.1~\ifb$ of $pp$ collisions at $\sqrt
s=13~\TeV$ recorded by the ATLAS detector at the LHC.  The results are
in agreement with SM predictions, and a substantial
region of the parameter space of a representative \zhdm model is excluded,
significantly improving upon previous results. Stringent limits are
also placed on the production cross-section of non-SM events with large
\met and a Higgs boson 
without extra model assumptions.

%%-------------------------------------------------------------------------------
%\section*{Acknowledgements}
%%-------------------------------------------------------------------------------

% Acknowledgements for papers with collision data
% Version 06-Mar-2017

% Standard acknowledgements start here
%----------------------------------------------
We thank CERN for the very successful operation of the LHC, as well as the
support staff from our institutions without whom ATLAS could not be
operated efficiently.

We acknowledge the support of ANPCyT, Argentina; YerPhI, Armenia; ARC, Australia; BMWFW and FWF, Austria; ANAS, Azerbaijan; SSTC, Belarus; CNPq and FAPESP, Brazil; NSERC, NRC and CFI, Canada; CERN; CONICYT, Chile; CAS, MOST and NSFC, China; COLCIENCIAS, Colombia; MSMT CR, MPO CR and VSC CR, Czech Republic; DNRF and DNSRC, Denmark; IN2P3-CNRS, CEA-DSM/IRFU, France; SRNSF, Georgia; BMBF, HGF, and MPG, Germany; GSRT, Greece; RGC, Hong Kong SAR, China; ISF, I-CORE and Benoziyo Center, Israel; INFN, Italy; MEXT and JSPS, Japan; CNRST, Morocco; NWO, Netherlands; RCN, Norway; MNiSW and NCN, Poland; FCT, Portugal; MNE/IFA, Romania; MES of Russia and NRC KI, Russian Federation; JINR; MESTD, Serbia; MSSR, Slovakia; ARRS and MIZ\v{S}, Slovenia; DST/NRF, South Africa; MINECO, Spain; SRC and Wallenberg Foundation, Sweden; SERI, SNSF and Cantons of Bern and Geneva, Switzerland; MOST, Taiwan; TAEK, Turkey; STFC, United Kingdom; DOE and NSF, United States of America. In addition, individual groups and members have received support from BCKDF, the Canada Council, CANARIE, CRC, Compute Canada, FQRNT, and the Ontario Innovation Trust, Canada; EPLANET, ERC, ERDF, FP7, Horizon 2020 and Marie Sk{\l}odowska-Curie Actions, European Union; Investissements d'Avenir Labex and Idex, ANR, R{\'e}gion Auvergne and Fondation Partager le Savoir, France; DFG and AvH Foundation, Germany; Herakleitos, Thales and Aristeia programmes co-financed by EU-ESF and the Greek NSRF; BSF, GIF and Minerva, Israel; BRF, Norway; CERCA Programme Generalitat de Catalunya, Generalitat Valenciana, Spain; the Royal Society and Leverhulme Trust, United Kingdom.

The crucial computing support from all WLCG partners is acknowledged gratefully, in particular from CERN, the ATLAS Tier-1 facilities at TRIUMF (Canada), NDGF (Denmark, Norway, Sweden), CC-IN2P3 (France), KIT/GridKA (Germany), INFN-CNAF (Italy), NL-T1 (Netherlands), PIC (Spain), ASGC (Taiwan), RAL (UK) and BNL (USA), the Tier-2 facilities worldwide and large non-WLCG resource providers. Major contributors of computing resources are listed in Ref.~\cite{ATL-GEN-PUB-2016-002}.
%----------------------------------------------

%The \texttt{atlaslatex} package contains the acknowledgements that were valid 
%at the time of the release you are using.
%These can be found in the \texttt{acknowledgements} subdirectory.
%When your ATLAS paper or PUB/CONF note is ready to be published,
%download the latest set of acknowledgements from:\\
%\url{https://twiki.cern.ch/twiki/bin/view/AtlasProtected/PubComAcknowledgements}

%-------------------------------------------------------------------------------
%\clearpage
%\appendix
%\part*{Appendix}
%\addcontentsline{toc}{part}{Appendix}
%-------------------------------------------------------------------------------

%In a paper, an appendix is used for technical details that would otherwise disturb the flow of the paper.
%Such an appendix should be printed before the Bibliography.

%%-------------------------------------------------------------------------------
%% If you use biblatex and either biber or bibtex to process the bibliography
%% just say \printbibliography here
\printbibliography
%% If you want to use the traditional BibTeX you need to use the syntax below.
%%\bibliographystyle{bibtex/bst/atlasBibStyleWoTitle}
%%\bibliography{MonoH_Paper_2016,bibtex/bib/ATLAS}
%%-------------------------------------------------------------------------------

\newpage

% ATLAS Collaboration author list
% Data extracted on 30-May-2017 for paper reference EXOT-2016-25
%\documentclass[11pt]{article}
%\usepackage{a4wide}\begin{document}
\begin{flushleft}
{\Large The ATLAS Collaboration}

\bigskip

M.~Aaboud$^\textrm{\scriptsize 137d}$,
G.~Aad$^\textrm{\scriptsize 88}$,
B.~Abbott$^\textrm{\scriptsize 115}$,
O.~Abdinov$^\textrm{\scriptsize 12}$$^{,*}$,
B.~Abeloos$^\textrm{\scriptsize 119}$,
S.H.~Abidi$^\textrm{\scriptsize 161}$,
O.S.~AbouZeid$^\textrm{\scriptsize 139}$,
N.L.~Abraham$^\textrm{\scriptsize 151}$,
H.~Abramowicz$^\textrm{\scriptsize 155}$,
H.~Abreu$^\textrm{\scriptsize 154}$
% ...
\bigskip
\\
% ...
$^{12}$ Institute of Physics, Azerbaijan Academy of Sciences, Baku, Azerbaijan\\
% ...
$^{88}$ CPPM, Aix-Marseille Universit{\'e} and CNRS/IN2P3, Marseille, France\\
% ...
$^{115}$ Homer L. Dodge Department of Physics and Astronomy, University of Oklahoma, Norman OK, United States of America\\
% ...
$^{119}$ LAL, Univ. Paris-Sud, CNRS/IN2P3, Universit{\'e} Paris-Saclay, Orsay, France\\
% ...
$^{137}$ $^{(a)}$ Facult{\'e} des Sciences Ain Chock, R{\'e}seau Universitaire de Physique des Hautes Energies - Universit{\'e} Hassan II, Casablanca; $^{(b)}$ Centre National de l'Energie des Sciences Techniques Nucleaires, Rabat; $^{(c)}$ Facult{\'e} des Sciences Semlalia, Universit{\'e} Cadi Ayyad, LPHEA-Marrakech; $^{(d)}$ Facult{\'e} des Sciences, Universit{\'e} Mohamed Premier and LPTPM, Oujda; $^{(e)}$ Facult{\'e} des sciences, Universit{\'e} Mohammed V, Rabat, Morocco\\
% ...
$^{139}$ Santa Cruz Institute for Particle Physics, University of California Santa Cruz, Santa Cruz CA, United States of America\\
% ...
$^{151}$ Department of Physics and Astronomy, University of Sussex, Brighton, United Kingdom\\
% ...
$^{154}$ Department of Physics, Technion: Israel Institute of Technology, Haifa, Israel\\
$^{155}$ Raymond and Beverly Sackler School of Physics and Astronomy, Tel Aviv University, Tel Aviv, Israel\\
% ...
$^{161}$ Department of Physics, University of Toronto, Toronto ON, Canada\\
% ...
$^{a}$ Also at Department of Physics, King's College London, London, United Kingdom\\
$^{b}$ Also at Institute of Physics, Azerbaijan Academy of Sciences, Baku, Azerbaijan\\
$^{c}$ Also at Novosibirsk State University, Novosibirsk, Russia\\
$^{d}$ Also at TRIUMF, Vancouver BC, Canada\\
$^{e}$ Also at Department of Physics {\&} Astronomy, University of Louisville, Louisville, KY, United States of America\\
% ...
$^{*}$ Deceased
\end{flushleft}

%\end{document}
% Created with xml2latex.py

\end{document}